\pdfoutput=1
\documentclass[a4paper,11pt]{article}
\usepackage[utf8]{inputenc}
\usepackage[table]{xcolor}
\usepackage{jcappub}
\usepackage{graphicx}
\usepackage{psfrag,fancyhdr,epsfig}
\usepackage{hyperref}
\hypersetup{colorlinks,bookmarksopen,bookmarksnumbered,citecolor=magenta,linkcolor=blue,pdfstartview=FitH,urlcolor=blue}
\usepackage{amsmath}
\usepackage{array}
\usepackage{graphicx}
\usepackage{color}
\usepackage{tensor}
\usepackage{xcolor}
\usepackage{orcidlink}
\usepackage{subcaption}
\newcommand{\be}{\begin{equation}}
\newcommand{\ee}{\end{equation}}
\newcommand{\bear}{\begin{array}}
\newcommand{\eear}{\end{array}}
\newcommand{\ba}{\begin{eqnarray}}
\newcommand{\ea}{\end{eqnarray}}

\def\a{\alpha}
\def\b{\beta}

\def\d{\delta}
\def\e{\epsilon}
\def\eps{\epsilon}
\def\g{\gamma}

\def\l{\lambda}
\def\m{\mu}
\def\n{\nu}

\def\r{\rho}
\def\s{\sigma}

\def\z{\zeta}

\def\tns{\tensor}
\def\cR{\mathcal{R}}

\usepackage[most]{tcolorbox}

\tcbset{colback=yellow!10!white, colframe=red!50!black, 
        highlight math style= {enhanced, %<-- needed for the ’remember’ options
            colframe=red,colback=red!10!white,boxsep=0pt}
        }

        \usepackage{empheq}
\newtcbox{\mymath}[1][]{%
    nobeforeafter, math upper, tcbox raise base,
    enhanced, colframe=blue!30!black,
    colback=blue!30, boxrule=1pt,
    #1}

\title{General Einstein-Cartan quadratic gravity with derivative couplings}


\author[a]{Theodoros Katsoulas\orcidlink{0000-0003-4103-7937}}
\author[a]{and Kyriakos Tamvakis\orcidlink{0009-0007-7953-9816}}
\emailAdd{th.katsoulas@uoi.gr} 
\emailAdd{tamvakis@uoi.gr}

\affiliation[a]{Physics Department, University of Ioannina, 45110, Ioannina, Greece}

\abstract{Within the framework of Einstein-Cartan gravity we consider an action, containing up to quadratic terms of the Ricci scalar and the Holst invariant, coupled non-minimally to a scalar field, including couplings of its derivatives to curvature. We derive the equivalent metric theory, featuring an extra dynamical pseudoscalar degree of freedom associated with the presence of the Holst term in the action. We study the evolution of the resulting two-field system in a FRW background and show that it evolves rapidly into an effective single-field inflationary model. We find that the model is consistent with the latest observational data for a wide range of its parameters, determining the necessary upper limits on derivative coupling parameters.}

\begin{document}

\maketitle

\section{Introduction} The standard framework of cosmological models is Einstein's theory of General Relativity (GR) with gravity entering in the action through the Einstein-Hilbert term of scalar curvature. An alternative formulation considers not only the metric but also the connection as an independent variable~\cite{Palatini1919, Hehl:1994ue}. This is the metric-affine formulation, which is equivalent to the standard metric formulation as long as we stay with the Einstein-Hilbert action, while the two formulations are not equivalent if we go beyond this action. Nevertheless, there are strong reasons to expect modifications. Gravitating matter fields are sensitive to quantum interactions that will ultimately introduce correction terms in the effective classical theory of gravity. Such terms are non-minimal couplings of matter fields to curvature like $\phi^2{\cal{R}}$, with their consequences studied extensively in the literature~\cite{Bezrukov:2007ep, Bauer:2008zj, Rasanen:2017ivk, Tenkanen:2017jih, Racioppi:2017spw, Markkanen:2017tun, Jarv:2017azx, Fu:2017iqg, Racioppi:2018zoy, Kozak:2018vlp, Rasanen:2018ihz, Almeida:2018oid, Shimada:2018lnm, Takahashi:2018brt, Jinno:2018jei, Rubio:2019ypq, Racioppi:2019jsp,Shaposhnikov:2020fdv, Borowiec:2020lfx, Jarv:2020qqm, Karam:2020rpa, McDonald:2020lpz, Langvik:2020nrs, Shaposhnikov:2020gts, Shaposhnikov:2020frq, Mikura:2020qhc, Verner:2020gfa, Enckell:2020lvn, Reyimuaji:2020goi, Karam:2021wzz,  Mikura:2021ldx, Racioppi:2021ynx,  Mikura:2021clt, Cheong:2021kyc, Azri:2021uat, Racioppi:2021jai,Piani:2022gon, Karananas:2022byw, Rigouzzo:2022yan, Gialamas:2022gxv,Hyun:2023bkf,Piani:2023aof,Gialamas:2023emn,Rigouzzo:2023sbb,Barman:2023opy}. In addition, coupling terms of the curvature to the derivatives of matter fields have been considered, like the terms $(\partial_{ \mu}\phi)(\partial_{ \nu}\phi)\,{\cal{R}}^{ \mu\nu}$, possibly arising from string theory~\cite{Gross:1986mw} or terms  $(\partial\phi)^2\cR$ that arise from radiative corrections in GR ~\cite{tHooft:1974toh}. The consequences of these terms have also been studied in the literature~\cite{Amendola:1993uh, Capozziello:1999uwa, Capozziello:1999xt, Germani:2010gm, Tsujikawa:2012mk,Kamada:2012se, Sadjadi:2012zp, Koutsoumbas:2013boa,Ema:2015oaa, Gumjudpai:2015vio, Zhu:2015lry, Sheikhahmadi:2016wyz,Dalianis:2016wpu,Harko:2016xip, Tumurtushaa:2019bmc, Fu:2019ttf,Dalianis:2019vit, Sato:2020ghj, Karydas:2021wmx}. In addition to these, higher powers of the curvature $\cR^2$~\cite{Davies:1977ze, Starobinsky:1980te}, attributed to quantum corrections of gravitating matter, have also been considered. Due to these modifications the predictions resulting from the metric-affine framework will be different than those obtained from the standard metric one~\cite{Meng:2004yf, Borunda:2008kf, Bombacigno:2018tyw, Enckell:2018hmo, Iosifidis:2018zjj, Antoniadis:2018ywb, Antoniadis:2018yfq, Tenkanen:2019jiq, Edery:2019txq, Giovannini:2019mgk, Edery:2019bsh, Gialamas:2019nly, Lloyd-Stubbs:2020pvx, Antoniadis:2020dfq,  Ghilencea:2020piz, Das:2020kff, Gialamas:2020snr, Ghilencea:2020rxc, Iosifidis:2020dck, Bekov:2020dww, Dimopoulos:2020pas,Karam:2021sno, Lykkas:2021vax, Gialamas:2021enw, Antoniadis:2021axu,  Gialamas:2021rpr, AlHallak:2021hwb,  Dioguardi:2021fmr,Dimopoulos:2022tvn, Dimopoulos:2022rdp, Pradisi:2022nmh, Durrer:2022emo, Salvio:2022suk, Antoniadis:2022cqh,  Lahanas:2022mng, Gialamas:2022xtt, Dioguardi:2022oqu,Iosifidis:2022xvp,Gialamas:2023aim, Gialamas:2023flv,SanchezLopez:2023ixx,Dioguardi:2023jwa,DiMarco:2023ncs,Gomes:2023xzk,Hu:2023yjn, Gialamas:2024iyu, Gialamas:2024jeb, Gialamas:2024uar} . Cosmological inflation~\cite{Kazanas:1980tx, Sato:1980yn, Guth:1980zm, Linde:1981mu} is presently accepted as the best explanation for the origin of the large scale structure of the Universe, providing the mechanism through which the primordial quantum fluctuations of gravitational and matter fields have been promoted to the cosmological perturbations presently observed~\cite{Starobinsky:1979ty, Mukhanov:1981xt, Hawking:1982cz, Starobinsky:1982ee, Guth:1982ec, Bardeen:1983qw}. Models of inflation are formulated in terms of a gravitating scalar field, the inflaton, which will in general participate in non-minimal corrections mentions above. The resulting inflationary predictions will be different, depending on whether the standard metric GR or the metric-affine framework are considered.

Ample motivation to consider formulations of gravity beyond the standard metric one is provided by presently existing tensions and questions of cosmology. In the metric-affine gravity the independent connection can be decomposed into the standard Levi-Civita part, arising in the metric case, and a part composed of the {\textit{torsion}} and {\textit{nonmetricity}} tensors. A special case of the general metric-affine framework is the case of Einstein-Cartan gravity~\cite{Cartan:1923zea, Cartan:1924yea}, characterized by non-zero torsion and zero nonmetricity. In this framework, apart from the Ricci scalar curvature ${\cal{R}}$, there is also a second linear scalar invariant of the curvature, namely the parity-odd Holst invariant $\tilde{\cal{R}}\sim \tns{\e}{_\a^\b^\g^\d}{\cal{R}}^{\alpha}_{\,\,\beta\gamma\delta}$. It is known that the presence of a quadratic Holst invariant term signals the presence of a dynamical pseudoscalar degree of freedom in the
spectrum~\cite{Pradisi:2022nmh, Gialamas:2022xtt}. A ghost-free action can be constructed containing at most quadratic terms of these invariants, including non-linear couplings to matter fields. Curvature couplings to derivatives of matter fields can also be written in terms of the Ricci tensor, like $(\partial_{ \mu}\phi)(\partial_{ \nu}\phi){\cal{R}}^{ \mu\nu}$~\cite{Gialamas:2020vto,Nezhad:2023dys, Gialamas:2024jeb}, and are also ghost-free.

In the present article we have considered Einstein-Cartan gravity characterized by non-zero torsion coupled to a scalar field with a quartic self-interaction potential. We employ the most general action quadratic in the curvature, devoid of unphysical degrees of freedom~\cite{BeltranJimenez:2019acz,BeltranJimenez:2020sqf,Marzo:2021iok,Annala:2022gtl,Barker:2024ydb,Barker:2024dhb}, consisting of at most quadratic terms of the Ricci scalar and the Holst invariant as well as non-minimal couplings of the scalar field and its derivatives to both. We transform the theory to the Einstein frame via a disformal transformation and express the resulting metric action in terms of the torsion. Integrating out the torsion the resulting theory describes a pair of scalar fields, namely the original fundamental scalar as well as a pseudoscalar field associated with the Holst invariant. We derive the set of equations of motion and analyze numerically the evolution of the two scalars in a FRW background. We find that the two scalar field system is found to evolves rapidly into an effective single-field system with a non-canonical kinetic term and a potential of the Palatini-$R^2$ form that possesses an inflationary plateau at large field values. We analyze the inflationary predictions of the model and in particular the effect of derivative couplings on the resulting values of inflationary observables. In section~\ref{Framework} we give a brief account of the general metric-affine framework of the theory to be considered. In section~\ref{Einstein_Cartan} we introduce the model, writing down the most general action of Einstein-Cartan gravity that includes up to quadratic curvature terms and is devoid of ghosts, coupled to a scalar field $\phi$ non-minimally, including couplings of the kinetic tensor $X_{ \mu\nu}=\partial_{ \mu}\phi\partial_{ \nu}\phi$ to curvature. We procced to transform the theory into the Einstein frame through a disformal transformation. In section~\ref{The_theory} we focus on the resulting $\mathcal{O}(X)$ theory, containing no more than quadratic kinetic terms of the scalar $\phi$ as well as the emerging dynamical pseudoscalar that results from the presence of the Holst term. In section~\ref{single_field} we analyze the resulting effective single-field theory. Section~\ref{inflation} is devoted in the study of the inflationary behavior of the model. Finally, in section~\ref{summary} we briefly summarize our results.
\section{Framework}
\label{Framework}
In the framework of Metric-Affine theory of gravity not only the metric $g_{ \mu\nu}$ but also the connection $\tns{\Gamma}{_\m^\r_\n}$ is an independent variable. The curvature is defined in terms of the connection as
\be
\tns{\cR}{^\a_\b_\g_\d} =\partial_\g \tns{\Gamma}{^\a_\d_\b} - \partial_\d \tns{\Gamma}{^\a_\g_\b} +\tns{\Gamma}{^\a_\g_\m}\tns{\Gamma}{^\m_\d_\b} -\tns{\Gamma}{^\a_\d_\m}\tns{\Gamma}{^\m_\g_\b}\,.\ee 
The only symmetry of $\tns{\cR}{^\a_\b_\g_\d}$  is the antisymmetry under the interchange of the last two indices. There are three distinct two-index contractions given by
\be
\tns{\cR}{_\m_\n} = \tns{\cR}{^\r_\m_\r_\n}\,, \qquad \tns{\widehat{\cR}}{^\m_\n} = g^{\a\b} \tns{\cR}{^\m_\a_\b_\n}\,, \qquad \tns{\tilde{\cR}}{_\m_\n} = \tns{\cR}{^\a_\a_\m_\n}\,,
\ee
called the Ricci, co-Ricci, and homothetic Ricci tensors. 
There are two possible curvature scalars derived from the Riemmann tensor, namely the Ricci scalar, determined through an additional contraction of either the Ricci tensor or the co-Ricci tensor,
\be
\cR = g^{\m\n}\tns{\cR}{_\m_\n} = -\tns{\widehat{\cR}}{^\m_\m}\,
\ee
and the Holst invariant

\be   
\tilde{\cal{R}}=\tns{\e}{_\m^\n^\r^\s}\tns{\cR}{^\m_\n_\r_\s}=g_{\m\a}\e^{\a\n\r\s}\tns{\cR}{^\m_\n_\r_\s}=\frac{1}{\sqrt{-g}}g_{\a\m}\varepsilon^{\a\n\r\s}\tns{\cR}{^\m_\n_\r_\s}.
\ee

The {\textit{torsion}} tensor $\tns{T}{^\r_\m_\n}$ is defined as
\be \tns{T}{^\r_\m_\n}\,=\,
\tns{\Gamma}{^\r_\m_\n}\,-\tns{\Gamma}{^\r_\n_\m}\,.\ee 
The torsion vanishes in the metric case, where the connection is fixed by the Levi-Civita relation\footnote{$\tns{\Gamma}{^\m_\n_\r}(g)=\frac{1}{2}g^{ \rho\sigma}\left(\partial_{ \mu}g_{ \sigma\nu}+\partial_{ \nu}g_{\mu\sigma}-\partial_{ \sigma}g_{ \mu\nu}\right)$} to be symmetric in its lower indices. In addition to the non-zero torsion the metric-affine framework is also characterized in general by the {\textit{non-metricity}} tensor, defined as
\be Q_{ \rho\mu\nu}\,=\,\nabla_{ \rho}g_{ \mu\nu}\,=\,\partial_{\rho}g_{ \mu\nu}-\tns{\Gamma}{^\s_\r_\m}g_{\s\n}-\tns{\Gamma}{^\s_\r_\n}g_{ \sigma\mu}\,.\ee
While the curvature measures the change of a vector under rotation after parallel transport in a closed loop, the torsion measures the non-closure of parallelograms formed from parallel -transported vectors. Finally, the non-metricity measures the change of the length of parallel-transported vectors. From the torsion tensor we may define a torsion vector $T_{ \mu}$ and an torsion axial vector $\hat{T}_{ \mu}$~\cite{Hehl:1994ue, Obukhov:1996pf, Obukhov:1997zd, Rigouzzo:2022yan} as
\be T_{ \mu}\,\equiv\,\tns{T}{^\nu_\mu_\nu},\,\,\,\,\,\,\,\hat{T}_{ \mu}=\tns{\e}{_\m^\n^\r^\s}T_{\n\r\s}\,.\ee
In terms of them we can write the torsion as
\be
T_{\a\b\g}=t_{\a\b\g}-\frac{1}{3}\Big(g_{\a\b}T_{\g}-g_{\a\g}T_{\b}\Big)+\frac{1}{6}\e_{\a\b\g\n}\hat{T}^\n\,,
\ee
where $t_{ \alpha\beta\gamma}$ is a purely tensorial part satisfying
\be 
g^{\m\n}t_{\m\r\n}=\e^{\r\s\m\n}t_{\s\m\n}=0\,.
\ee
Similarly, we may define the non-metricity vectors $Q_{ \mu}$ and $\hat{Q}_{ \mu}$ as
\be Q_{ \mu}\,\equiv\,\tns{Q}{_\m^\n_\n}\,\,,\,\,\,\,\,\,\,\hat{Q}_{ \mu}\,\equiv\,\tns{Q}{^\n_\n_\m}\,.\ee
The non-metricity tensor can be written in terms of them as
\be 
Q_{\a\b\g}=q_{\a\b\g}+\frac{1}{18}\Big[g_{\b\g}(5Q_\a-2\hat{Q}_\a)+4g_{\a\b}\hat{Q}_\g+4g_{\a\g}\hat{Q}_\b -g_{\a\b}Q_{\g}-g_{\a\g}Q_{\b}\Big]\,,
\ee
where $q_{\a\b\g}$ is a purely tensorial part satisfying 
\be 
g^{\a\b}q_{\g\a\b}=g^{\a\b}q_{\a\g\b}=0\,\,.
\ee
The connection $\Gamma^{\alpha}_{\,\beta\gamma}$ can be decomposed in terms of these vectors as
\be\begin{split}
\tns{\Gamma}{^\a_\b_\g}&=\tns{\Gamma}{^\a_\b_\g}(g)\,+\,\frac{1}{2}\Bigg(-\tns{\d}{^\a_\b}\Big(\frac{1}{18}\Big(5Q_{ \gamma}-2\hat{Q}_{\gamma}\Big)+\frac{2}{3}T_{\gamma}\Big)+g_{ \beta\gamma}\Big(\frac{1}{18}\Big(7Q^{\alpha}-10\hat{Q}^{\alpha}\Big)+\frac{2}{3}T^{ \alpha}\Big)\\&
-\tns{\d}{^\a_\g}\frac{1}{18}\left(5Q_{\beta}-2\hat{Q}_{\beta}\right)+\frac{1}{6}\tns{\e}{^\a_\b_\g_\d}\hat{T}^{ \delta}+\tau_{\beta\,\,\,\,\gamma}^{\,\,\,\alpha}\Bigg),
\end{split}
\ee
where $\tns{\Gamma}{^\a_\b_\g}(g)$ is the Levi-Civita connection and
\be
\tau_{\b\a\g}=q_{\a\b\g}-q_{\g\a\b}-q_{\b\a\g}+t_{\a\b\g}+t_{\g\a\b}+t_{\b\a\g}
\ee
is the overall purely tensorial part. Substituting the above expression of the connection into the Riemann tensor we obtain an expression of it in terms of the metric Riemann tensor $\tns{R}{^\a_\b_\g_\d}(g) $ and $T,\,\hat{T},\,Q,\,\hat{Q},\,\tau$'s.

In the case of the Einstein-Cartan theory of gravity, characterized only by non-zero torsion, $Q,\,q$ are set to zero. We also set the tensorial part $t_{\alpha\beta\gamma}$ to zero, avoiding the occurence of unphysical degrees of freedom associated with it~\cite{Barker:2024ydb,Barker:2024juc}. In this case the Riemann tensor is
\be 
\begin{split}
    \tns{\cR}{^\a_\b_\g_\d}&=\tns{R}{^\a_\b_\g_\d}(g)-\frac{1}{3}\tns{\d}{^\a_\d}\nabla_\g T_\b+\frac{1}{3}g_{\b\d}\nabla_\g T^\a-\frac{1}{12}\tns{\e}{_\d^\a_\b_\n}\nabla_\g\hat{T}^\n+\frac{1}{3}\tns{\d}{^\a_\g}\nabla_\d T_\b-\frac{1}{3}g_{\b\g}\nabla_\d T^\a\\&+\frac{1}{12}\tns{\e}{_\g^\a_\b_\n}\nabla_\d\hat{T}^\n+\frac{1}{9}\tns{\d}{^\a_\g}T_\b T_\d-\frac{1}{9}g_{\b\d}\tns{\d}{^\a_\g}T_\mu T^\mu+\frac{1}{36}\tns{\e}{_\d^\m_\b_\l}\tns{\d}{^\a_\g}T_\m\hat{T}^\l-\frac{1}{9}g_{\g\d}T^\a T_\b\\&+\frac{1}{9}g_{\b\d}T^\a T_{\g}-\frac{1}{36}\tns{\e}{_\d_\g_\b_\l}T^\a\hat{T}^\l+\frac{1}{36}\tns{\e}{_\g^\a_\d_\n}T_\b \hat{T}^\n-\frac{1}{36}\tns{\e}{_\g^\a_\m_\n}g_{\b\d}T^\m\hat{T}^\n-\frac{1}{9}\tns{\d}{^\a_\d}T_{\b}T_{\g}\\&+\frac{1}{9}g_{\b\g}\tns{\d}{^\a_\d}T_\mu T^\mu-\frac{1}{36}\tns{\e}{_\g^\m_\b_\l}\tns{\d}{^\a_\d}T_{\m}\hat{T}^\l+\frac{1}{9}g_{\g\d}T^\a T_{\b}-\frac{1}{9}g_{\b\g}T^\a T_\d +\frac{1}{36}\tns{\e}{_\g_\d_\b_\l}T^\a\hat{T}^\l\\&-\frac{1}{36}\tns{\e}{_\d^\a_\g_\n}T_\b\hat{T}^\n+\frac{1}{36}\tns{\e}{_\d^\a_\m_\n}g_{\b\g}T^\m\hat{T}^\n-\frac{1}{144}\tns{\e}{_\d^\a_\m_\n}\tns{\e}{_\g^\m_\b_\l}\hat{T}^\n\hat{T}^\l+\frac{1}{144}\tns{\e}{_\g^\a_\m_\n}\tns{\e}{_\d^\m_\b_\l}\hat{T}^\n\hat{T}^\l.
\end{split}{\label{RIEMANN}}
\ee
\section{Einstein-Cartan quadratic gravity with derivative couplings}
\label{Einstein_Cartan}
As we discussed in the introduction the quantum interactions of gravitating matter fields are bound to generate corrections to the Hilbert-Einstein gravitational action. These modifications can be non-minimal couplings of matter fields and their derivatives to curvature, like $f(\phi){\cal{R}}$, $g(\phi)\tilde{\cal{R}}$, $(\partial^{ \mu}\phi)(\partial^{ \nu}\phi){\cal{R}}_{ \mu\nu}$,
or higher powers of the curvature, like ${\cal{R}}^2$, $\tilde{\cal{R}}^2$. We proceed to write down an effective theory of gravity coupled to a scalar field that is at most quadratic in powers of the curvature and ghost free. Therefore, we consider the action (in reduced Planck mass units)
\be
\begin{split}
{\cal{S}}\,=\,\int\,{\rm d}^4x\,\sqrt{-g}\,\Bigg\{\,&\frac{1}{2}f(\phi){\cal{R}}\,+\,\frac{1}{2}g(\phi)\tilde{\cal{R}}\,+\,\frac{1}{2}\alpha_1X{\cal{R}}\,-\frac{1}{2}K(\phi)X-V(\phi)\\&
+\alpha_2{\cal{R}}^{ \mu\nu}X_{ \mu\nu}\,+\,\alpha_3\hat{\mathcal{R}}^{\mu\nu}X_{ \mu\nu}\, \,+\,\frac{\beta}{4}{\cal{R}}^2\,+\,\frac{\delta}{4}\tilde{\cal{R}}^2  \Bigg\}
\end{split} {\label{ACT-0}}
\ee
where $X_{\mu\nu}=(\partial_{ \mu}\phi)(\partial_{ \nu}\phi)$ and $X=(\partial\phi)^2$.
An extra term $\alpha_4\,X\tilde{\cal{R}}$ is also possible under our assumptions but we ignore it, since it corresponds to a hard breaking of parity. The action ({\ref{ACT-0}}) can be rewritten in terms of two auxiliary fields $\chi$ and $\zeta$ as\footnote{A parity-odd mixing term ${\cal{R}}\tilde{\cal{R}}$~\cite{Gialamas:2024iyu} in ({\ref{ACT-0}}) would simply modify the potential by a mixing term of the auxiliaries $\chi\zeta$ and the analysis to follow would be very much analogous. So, for the sake of simplicity such a term is not included.}
\be \begin{split}
{\cal{S}}\,=\,\int\,{\rm d}^4x\,\sqrt{-g}\,\Bigg\{\,&\frac{1}{2}\left({\cal{F}}(\phi,\,\chi)+\alpha_1 X\right){\cal{R}}\,+\,\frac{1}{2}{\cal{G}}(\phi,\,\zeta)\tilde{\cal{R}}\\&
-\frac{1}{2}K(\phi)X\,-U(\phi,\,\chi,\,\zeta)\,+\,\alpha_2{\cal{R}}^{ \mu\nu}X_{ \mu\nu}\,+\,\alpha_3\overline{\cR}^{ \mu\nu}X_{ \mu\nu}\,\Bigg\},
\end{split}   
{\label{ACT-1}}  
\ee
where
\be\begin{split}
{\cal{F}}(\phi,\chi)=f(\phi)+\beta\chi\,,\,\,\,\,\,\,\,\,{\cal{G}}(\phi,\,\zeta)\,=\,g(\phi)+\delta\zeta\\
\,\\
U(\phi,\,\chi,\,\zeta)\,=\,V(\phi)+\frac{\beta}{4}\chi^2\,+\,\frac{\delta}{4}\zeta^2.
\end{split}
\ee
We have replaced the coupling to the co-Ricci tensor in ({\ref{ACT-0}}) with a coupling to the so-called average Ricci tensor $\overline{\mathcal{R}}_{ \mu\nu}=\left(\mathcal{R}_{ \mu\nu}+\hat{\mathcal{R}}_{ \mu\nu}\right)/2$, which vanishes if the theory is metric compatible. Therefore, focusing on the Einstein-Cartan framework of gravity, we may eliminate the coupling $\alpha_3$ in ({\ref{ACT-1}}).

In order to transform the action to the Einstein frame we need to employ a {\textit{disformal transformation}}~\cite{Nezhad:2023dys} $g_{ \mu\nu}\rightarrow\,\tilde{g}_{ \mu\nu}$, defined by\footnote{Note that invertibility of the disformal transformation requires $\gamma_1>0$, $\gamma_2>0$ and $\tilde{\gamma}_1-X\partial\tilde{\gamma}_1/\partial X-X^2\partial\tilde{\gamma}_2/\partial X\,\neq\,0$.}
\be g_{ \mu\nu}=\gamma_1(\phi,\,\tilde{X})\tilde{g}_{ \mu\nu}+\gamma_2(\phi,\tilde{X})X_{ \mu\nu},\,\,\,\,\,\,\,\,\,\,\,\tilde{g}_{ \mu\nu}=\tilde{\gamma}_1(\phi,X)g_{ \mu\nu}+\tilde{\gamma}_2(\phi,X)X_{ \mu\nu}\,,\ee
parametrized by the functions $\gamma_1$ and $\gamma_2$. Note that $\tilde{\gamma}_1=\gamma_1^{-1},\,\tilde{\gamma}_2=-\gamma_2/\gamma_1$ and $X=\tilde{X}\left(\gamma_1(\phi,\tilde{X})+\gamma_2(\phi,\tilde{X})\tilde{X}\right)^{-1}$ and $\tilde{X}=X\left(\tilde{\gamma}_1(\phi,X)+\tilde{\gamma}_2(\phi,X)X\right)^{-1}$.

Applying the disformal transformation on the action ({\ref{ACT-1}}) we obtain\footnote{The disformal transformation of the Hols invariant is
$$\sqrt{-g}\tilde{\cal{R}}=\sqrt{-\tilde{g}}\left(\gamma_1(\phi,\tilde{X})\tilde{g}_{ \mu\nu}+\gamma_2(\phi,\tilde{X})X_{ \mu\nu}\right)\tilde{\epsilon}^{ \mu\beta\gamma\delta}{\cal{R}}^{ \alpha}_{\,\,\beta\gamma\delta}$$}
\be \begin{split}
{\cal{S}}=\int\,{\rm d}^4x\,\sqrt{-g}\Bigg\{\,&\frac{1}{2}F_1{\cal{R}}+F_2\overline{\cal{R}}^{ \mu\nu}X_{ \mu\nu}+F_3{\cal{R}}^{ \mu\nu}X_{ \mu\nu}\\&
+F_4\tilde{\cal{R}}+F_5X_{ \alpha\mu}\epsilon^{ \mu\beta\gamma\delta}{\cal{R}}^{ \alpha}_{\,\,\beta\gamma\delta}-F_6X-F_7U\Bigg\}.
\end{split}
\ee
For simplicity of notation we have omitted the tildes. The coefficient functions are ($\gamma\equiv \frac{\gamma_2}{\gamma_1}$)
\be
\begin{split}
&F_1(\phi,X,\chi)=(1+\g X)^{1/2}\Big(\g_1\mathcal{F}(\phi,X)+\frac{\a_1 X}{1+\g X}\Big)\\&
F_2(\phi,X)=\a_3(1+\g X)^{-1/2}\\&
F_3(\phi,X,\chi)=\frac{1}{2}(1+\g X)^{-1/2}\Big(-\g_2\mathcal{F}(\phi,X)+\frac{\a_2-\a_3-(\a_1+\a_3)\g X}{1+\g X}\Big)\\&
F_4(\phi,\z,X)=\frac{1}{2}\g_1\mathcal{G}(\phi,\z)=\frac{F_5(\phi,\z)}{\g}\\&
F_6(\phi,X)=\frac{1}{2}\g_1(1+\g X)^{-1/2}K(\phi)\\&
F_7(\phi,X,\chi)=\g_1^2(1+\g X)^{1/2}
\end{split}
\ee
From now on we shall focus on the case of Einstein-Cartan gravity. This corresponds to setting $\alpha_3$ to zero. The Einstein frame is defined by the two conditions
\be\begin{split}
&F_1(\phi,X,\chi)=(1+\g X)^{1/2}\Big(\g_1\mathcal{F}(\phi,X)+\frac{\a_1 X}{1+\g X}\Big)\,=\,1\\&
F_3(\phi,X,\chi)=\frac{1}{2}(1+\g X)^{-1/2}\Big(-\g_2\mathcal{F}(\phi,X)+\frac{\a_2-\a_3-(\a_1+\a_3)\g X}{1+\g X}\Big)\,=\,0\,.
\end{split}{\label{COND}}
\ee
The conditions ({\ref{COND}}) determine the parametric functions of the disformal transformation. Assuming that the kinetic function is not too large, we may obtain to $O(X)$ 
\be
\g \simeq \a_2 -\frac{\a_2^2}{2}X\,, \quad \g_1\simeq \frac{1}{\mathcal{F}(\phi,\chi)}\left( 1-(\a_1+\a_2/2)X\right)\,,
\ee

The resulting Einstein frame action now is
\be 
{\cal{S}}=\int\,{\rm d}^4x\,\sqrt{-g}\Bigg\{\,\frac{1}{2}{\cal{R}}
+F_4\tilde{\cal{R}}+F_5X_{ \alpha\mu}\epsilon^{ \mu\beta\gamma\delta}{\cal{R}}^{ \alpha}_{\,\,\beta\gamma\delta}-F_6X-F_7U\Bigg\}.{\label{ACT-2}}
\ee
Next, we can substitute in ({\ref{ACT-2}}) the expression of the curvature in terms of the torsion ({\ref{RIEMANN}}) as well as the curvature scalars
\be \begin{split}
&{\cal{R}}=R(g)+2\nabla_{ \mu}T^{ \mu}-\frac{2}{3}T_{ \mu}T^{ \mu}+\frac{1}{24}\hat{T}_{ \mu}\hat{T}^{ \mu}\\&
\tilde{\cal{R}}=\frac{2}{3}T_{ \mu}\hat{T}^{ \mu}-\nabla_{ \mu}\hat{T}^{ \mu}
\end{split}\ee
where $R[g]$ is the standard metric Ricci scalar in terms of the Levi-Civita connection.
We obtain the Einstein frame action in a standard metric form as
\be
\begin{split}
    \mathcal{S}&=\int {\rm d}^4x\sqrt{-g}\Bigg\{\frac{1}{2}R-\frac{1}{3}T^{\mu}T_{\mu}+\frac{1}{48}\hat{T}^{\mu}\hat{T}_{\mu}+\frac{2}{3}F_4T\cdot\hat{T}+(\nabla F_4)\cdot\hat{T}+\frac{1}{3}\nabla_\m(F_5X)\hat{T}^\m\\&-\frac{1}{3}\nabla_\m(\tns{X}{^\m_\n}F_5)\hat{T}^\n+\frac{2}{9}F_5X_{\m\n}T^\m\hat{T}^\n+\frac{1}{9}F_5XT\cdot\hat{T}-\frac{1}{2}\g_1(1+\g X)^{-1/2}K(\phi)X\\&-\g_1^2(1+\g X)^{1/2}U(\phi,\chi,\z)  \Bigg\}.
\end{split} {\label{ACT-3}}
\ee
Varying with respect to the torsion vectors we obtain the algebraic system of equations
 \be 
\begin{split}
&T_\m=\Big(F_4+\frac{1}{6}F_5X\Big)\hat{T}_\m+\frac{1}{3}F_5X_{\m\n}\hat{T}^\n\\\\&
\frac{1}{16}\hat{T}_\m+\Big(F_4+\frac{1}{6}F_5X\Big)T_\m+\frac{1}{3}F_5X_{\m\n}T^\n=E_\m\,,
\end{split}
\ee
where  $E_\m$ is
\be E_{\mu}=-\frac{3}{2}\nabla_{\mu}(F_4-\frac{X}{3}F_5)-\frac{1}{2}\nabla^{\nu}(F_5X_{\mu\nu})\,.{\label{EPSILON}}\ee
The corresponding solutions for $T_{ \mu},\,\hat{T}_{\mu}$ are given in the Appendix. Substituting back into the action ({\ref{ACT-3}}), which is given in ({\ref{ACT-4}}) of the Appendix.

The starting action contained three scalar field variables, namrely $\phi,\,\chi,\,\zeta$. In the final action ({\ref{ACT-4}}) the appearing scalar variables are $\phi,\,\chi,\,\zeta,\,F_4,\,F_5$. Only three of them are independent due to the relations
\be F_4=\frac{1}{2}\left(g(\phi)+\delta\zeta\right)\gamma_1,\,\,F_5=\frac{1}{2}\left(g(\phi)+\delta\zeta\right)\gamma_2=F_4\frac{\gamma_2}{\gamma_1}\,.\ee
$\zeta$ can be replaced by $F_4$ and $\phi$ through the relation\footnote{The $O(X)$ expressions of the $\gamma$'s are
$$\gamma_1\approx\,\left(f(\phi)+\beta\chi\right)^{-1}\left(1-(\alpha_1+\alpha_2/2)X\right),\,\,\,\,\gamma\equiv \frac{\gamma_2}{\gamma_1}\approx\alpha_2\left(1-\frac{\alpha_2}{2}X\right)$$}
\be \zeta\,=\,\delta^{-1}\left(-g(\phi)+\frac{2}{\gamma_1}F_4\right)\,.\ee
Out of the three independent scalars only $F_4$ and $\phi$ possess kinetic terms, while $\chi$ is an auxiliary field that can be replaced by substituting into the action its algebraic equation of motion. Note that the dynamical scalar $F_4$, which from now on we shall denote with the symbol $\sigma$, was not a part of the starting matter action, which contained only $\phi$, but is of gravitational origin, arising from the presence of the Holst quadratic term in the action.

\section{The ${\cal{O}}(X)$ theory}
\label{The_theory}
In order to simplify notation we shall denote $F_4=\sigma$ (then, $F_5=\gamma\sigma$). The action, shown in ({\ref{ACT-7}}) of the Appendix
 consists of a purely kinetic piece, depending only on $\sigma$ and $\phi$, namely  
 \be\begin{split}
{\cal{S}}_I&=\int\,{\rm d}^4x\,\sqrt{-g}\Bigg\{\frac{-12+64(\alpha_2X-3)\sigma^2}{(1+16\sigma^2)^2}(\nabla_\m\sigma)^2-\frac{384\sigma^2}{(1+16\sigma^2)^2}X^{ \mu\nu}\nabla_{ \mu}\sigma\nabla_{ \nu}\sigma\\\\&
+\frac{\left(-8+\sigma+32\sigma^2(-20+\sigma+8\sigma^2\right)\alpha_2}{(1+16\sigma^2)^2}\Big(\nabla_{ \mu}\sigma\nabla^{ \mu}(\sigma X)-\nabla_{ \mu}(X^{ \mu\nu}\sigma)\nabla_{ \nu}\sigma\Big)\,\,+\,O(X^2)\,\Bigg\}
\end{split}\ee
and of a part consisting of all three scalars, namely 
\be
{\cal{S}}_{II}=\int\,{\rm d}^4x\,\sqrt{-g}\Bigg\{-\frac{1}{2}\g_1(1+\g X)^{-1/2}K(\phi)X-\g_1^2(1+\g X)^{1/2}U \Bigg\}
\ee
or
\be\begin{split}
\mathcal{S}_{II}=
\int\,{\rm d}^4x\,\sqrt{-g}\Bigg\{&-\frac{1}{2}\frac{\left(1-(\alpha_1+\alpha_2/2)X\right)}{\left(f(\phi)+\beta\chi\right)}(1+\alpha_2X)^{-1/2}K(\phi)X\\&-\frac{\left(1-(\alpha_1+\alpha_2/2)X\right)^2}{\left(f(\phi)+\beta\chi\right)^2}(1+\alpha_2X)^{1/2}U(\phi,\,\sigma,\,\chi)\Bigg\}\,,
\end{split}
\ee
where
\be U(\phi,\sigma,\,\chi)=V(\phi)+\frac{\beta}{4}\chi^2+\frac{1}{4\delta}\left(\frac{2\sigma\left(f(\phi)+\beta\chi\right)}{\left(1-(\alpha_1+\alpha_2/2)X\right)}-g(\phi)\right)^2\ee

At this point we can consider the equation of motion for the auxiliary field $\frac{\delta{\cal{S}}}{\delta\chi}=0$ having in mind that we seek for an $\mathcal{O}(X)$ solution. We obtain 
\be\begin{split}
&0=-\frac{\beta}{2}f(\phi)\chi-\frac{\beta}{\delta}\sigma g(\phi)(f(\phi)+\beta\chi)+2\beta V+\frac{\beta}{2\delta}g^2(\phi)\\&
+X\Bigg\{\frac{\beta}{2}K(\phi)(f(\phi)+\beta\chi)-\frac{\beta\hat{\alpha}'}{\delta}\sigma g(\phi)(f(\phi)+\beta\chi)+\frac{\hat{\alpha}\beta}{2}\chi f(\phi)\\&+\frac{\hat{\alpha}\beta}{\delta}\sigma (f(\phi)+\beta\chi)g(\phi)-2\beta\hat{\alpha}V(\phi)-\frac{\hat{\alpha}\beta}{2\delta}g^2(\phi)\Bigg\}.
\end{split}{\label{X-EQ}}
\ee
The zeroth-order solution is
\be \chi_0=\frac{4V(\phi)-\frac{2\sigma}{\delta}f(\phi)g(\phi)+\frac{g^2(\phi)}{\delta}}{f(\phi)+\frac{2\beta \sigma}{\delta}g)(\phi)}{\label{X0}}.\ee
Next, we insert in ({\ref{X-EQ}}) a perturbative Ansatz $\chi\,=\,\chi_0\,+\,X\,\chi_1$ and obtain the correction ($\hat{\alpha}'=\alpha_1+\alpha_2/2$)
\be \chi_1\,=\,\frac{\left(f^2(\phi)+4\beta V+\frac{\beta}{\delta}g^2(\phi)\right)\left(K(\phi)-2\hat{\alpha}'\frac{\beta}{\delta}\sigma g(\phi)\right)}{\left(f(\phi)+2\frac{\beta}{\delta}\sigma g(\phi)\right)^2}\,{\label{X1}}\,.\ee
{{Substituting the solution for $\chi$ to the $\mathcal{S}_{II}$, given in (\ref{ACT-6}), and combining the result with the kinetic action ${\cal{S}}_I$, we arrive at the complete $\mathcal{O}(X)$ action 
 }}
\be 
\begin{split}
    \mathcal{S}=\int {\rm d}^4x\sqrt{-g}\Bigg\{&\frac{1}{2}R-\frac{1}{2}\mathcal{K}_1(\phi,\s)X-\frac{1}{2}\mathcal{K}_2(\s,X)(\nabla_\m\s)^2-\mathcal{K}_3(\s)X^{\m\n}\nabla_{\m}\s\nabla_\n\s+\mathcal{K}_4(\s)\nabla_\m\s\nabla^\m(\s X)\\&-\mathcal{K}_5(\s)\nabla_\m(\s X^{\m\n})\nabla_\n\s-\tilde{U}(\phi,\s)\Bigg\}\,,
    \end{split}
\ee
\begin{figure}[t!]
    \centering
    \includegraphics[width=0.5\linewidth]{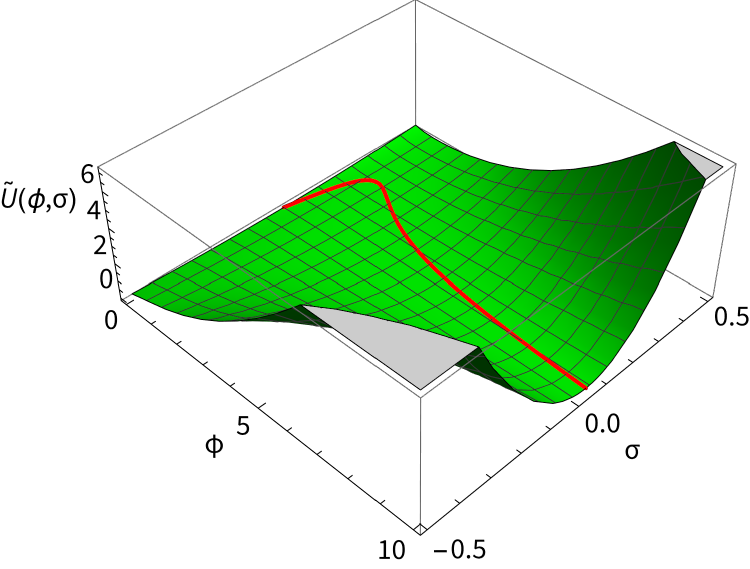}
    \caption{The potential $\tilde{U}(\phi,\,\sigma)$ for $f(\phi)=1+\xi\phi ^2,g(\phi)= \bar{\xi}\phi +\bar{\xi}'\phi^3, V(\phi)= \lambda/4\phi^4$, $\beta=10^6,\delta=\xi=\bar{\xi}= 1,\bar{\xi}'=0,\lambda=2.3\cdot10^{-10}$. }
    \label{fig:effective_potential}
\end{figure}
where 
\be
\begin{split}
    &\mathcal{K}_1(\phi,\s)=-\frac{1}{2(g^2(\phi)\b+f^2(\phi)\d+4\b \d V(\phi))}\Bigg(g^2(\phi)\hat{\a}-2f(\phi)K(\phi)-4g(\phi)((\hat{\a}-\hat{\a}')f(\phi)+\b K(\phi))\s\\&+4f^2(\phi)(\hat{\a}-2\hat{\alpha}')\s^2+4(\hat{\a}\d+4(\hat{\a}-2\hat{\a}')\b\s^2)V(\phi)\Bigg)\\\\& 
    \mathcal{K}_2(\s,X)=\frac{24-128(\a_2X-3)\s^2}{(1+16\s^2)^2},\,\,\,\,\,\,\,\,
    \mathcal{K}_3(\s)=\frac{384\s^2}{(1+16\s^2)^2}\\\\&
    \mathcal{K}_4(\s)=\frac{(-8+\s+32\s^2(-20+\s+8\s^2))\a_2}{(1+16\s^2)^2}=\mathcal{K}_5(\s)\\\\&
   \tilde{U}(\phi,\s)=\left(\frac{f^2(\phi)+4\beta V(\phi)}{\delta f^2(\phi)+\beta g^2(\phi)+4\beta\delta V(\phi)}\right)\left(\sigma-\frac{f(\phi)g(\phi)}{2(f^2(\phi)+4\beta V(\phi))}\right)^2+\frac{V(\phi)}{f^2(\phi)+4\beta V(\phi)} 
\end{split} {\label{ACT-FINAL}}
\ee
\section{The single-field model}
\label{single_field}
Having in mind to consider the inflationary behavior of the model at hand in the framework of slow-roll inflation, we may neglect acceleration terms as well as terms of higher that two powers of velocities. Doing that, we obtain the following equations of motion for the two scalars 
\be 
\begin{split}
    &\frac{1}{\sqrt{-g}}\partial_{\m}\Big(\sqrt{-g}\mathcal{K}_1(\phi,\s)\partial^\m\phi\Big)-\frac{1}{2}\frac{\partial \mathcal{K}_1(\phi,\s)}{\partial\phi}(\nabla_\m\phi)^2-\frac{\partial \tilde{U}(\phi,\s)}{\partial \phi}=0\\\\&
\frac{1}{\sqrt{-g}}\partial_{\m}\Big(\sqrt{-g}\mathcal{K}_2(\s)\partial^\m\s\Big)-\frac{1}{2}\frac{\partial\mathcal{K}_2(\s)}{\partial\s}(\nabla_\m\s)^2-\frac{1}{2}\frac{\partial\mathcal{K}_1(\phi,\s)}{\partial\s}(\nabla_\m\phi)^2-\frac{\partial \tilde{U}(\phi,\s)}{\partial\s}=0.
\end{split}{\label{EQS-0}}
\ee
The potential of the model, written in the form ({\ref{ACT-FINAL}}), has a manifestly minimum line in the two-field space along the direction
\be \sigma_0(\phi)\,=\,\frac{f(\phi)g(\phi)}{2\left(f^2(\phi)+4\beta V(\phi)\right)} {\label{MIN}}
\ee
Along this line it reduces to the potential encountered in the Palatini-${\cal{R}}^2$ models~\cite{Enckell:2018hmo, Antoniadis:2018ywb}~\cite{Gialamas:2022xtt}, which is characterized by an inflationary plateau at large values of $V(\phi)$. The potential $\tilde{U}(\phi,\,\sigma)$ and the corresponding minimum valley are shown in Fig~\ref{fig:effective_potential}. 

\begin{figure}[t!]
  \centering
  \subfloat[]{\includegraphics[width=0.48\textwidth]{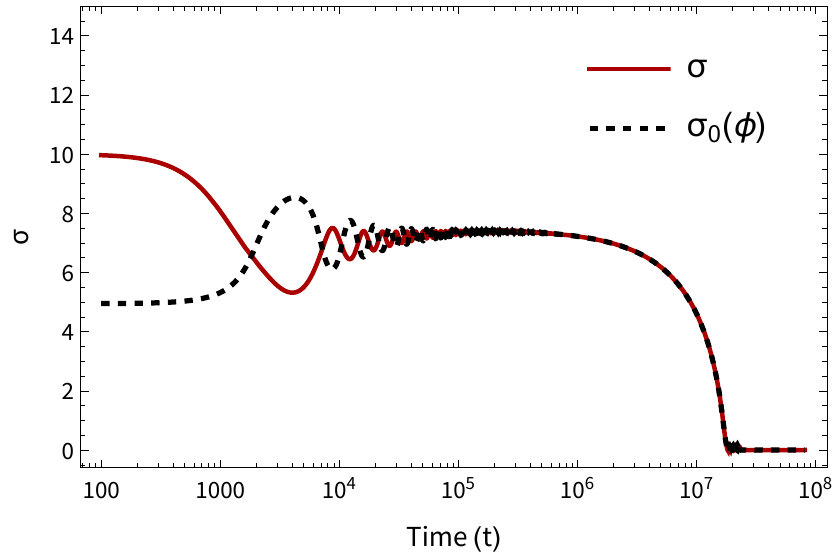}}
  \hfill
  \subfloat[]{\includegraphics[width=0.48\textwidth]{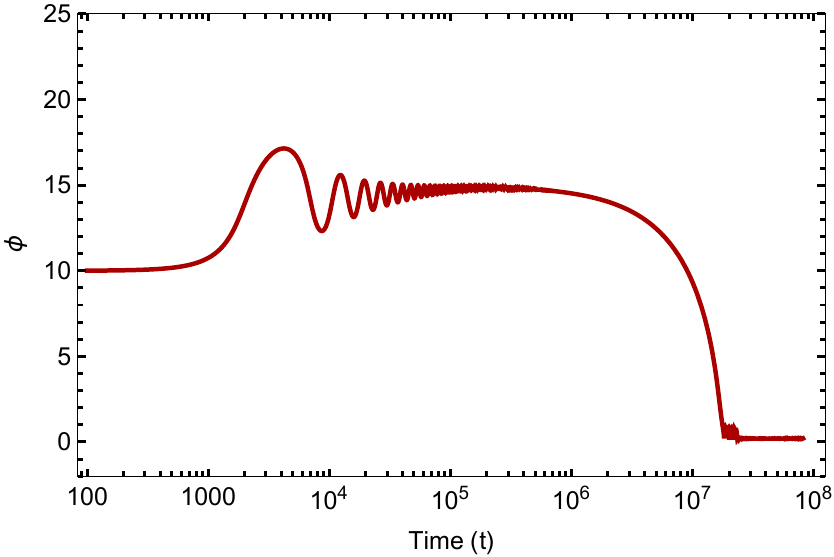}}
  \caption{(a)The pseudoscalar $\s$ and (b) the scalar field $\phi$ as functions of time $(t)$, for $f(\phi)=1+\xi\phi^2$, $g(\phi)=\bar{\xi}\phi+\bar{\xi}'\phi^3$, $V(\phi)=\l\phi^4/4$, $K(\phi)=1$, $\b=10^6$, $\xi=\d=1=\bar{\xi}'=1$, $\bar{\xi}=0$, $\hat{\a}=-10^5,\, \hat{\a}'=-10^2$. We have chosen the set of the parameters based on ~\cite{Gialamas:2022xtt, Gialamas:2024jeb}. We have checked that this general behavior remains for different values of the parameters $\hat\alpha$ and $\hat\alpha'$.}
  \label{PHI-SIGMA}
  \end{figure}

In order to discuss the inflationary behavior of the model at hand we consider an FRW background metric and write the resulting pair of scalar equations. They are
 \be 
\begin{split}
    &3H\mathcal{K}_1(\phi,\s)\dot{\phi}+\frac{1}{2}\frac{\partial\mathcal{K}_1(\phi,\s)}{\partial\phi}\dot{\phi}^2+\frac{\partial\mathcal{K}_1(\phi,\s)}{\partial \s}\dot{\phi}\dot{\s}+\mathcal{K}_1(\phi,\s)\ddot{\phi}+\frac{\partial \tilde{U}(\phi,\s)}{\partial\phi}=0\\\\&
    \mathcal{K}_2(\s)\ddot{\s}+3H\mathcal{K}_2\dot{\s}+\frac{1}{2}\frac{\partial \mathcal{K}_2(\s)}{\partial\s}\dot{\s}^2-\frac{1}{2}\frac{\partial \mathcal{K}_1(\phi,\s)}{\partial\s}\dot{\phi}^2+\frac{\partial \tilde{U}(\phi,\s)}{\partial\sigma}=0.
\end{split} \label{EQS-1}
\ee
The corresponding Friedmann equation is
\be H^2=\frac{\rho}{3}\,\,\,\,\,\,\,\,\,\,\,{\textit{with}}\,\,\,\,\,\,\,\,
        \rho=\frac{1}{2}\mathcal{K}_1(\phi,\s)\dot{\phi}^2+\frac{1}{2}\mathcal{K}_2(\s)\dot{\s}^2+\tilde{U}(\phi,\s)\,.\ee
We have solved numerically the coupled system of the equations ({\ref{EQS-1}}) for characteristic values of the parameters and initial conditions $\phi(0)=\sigma(0)=10$ and $\dot\phi(0)=\dot\s(0)=0$. The result is shown in Fig.~\ref{PHI-SIGMA}, justifying a rather fast evolution of the two-field system towards the minimum line ({\ref{MIN}}). Therefore, we can safely conclude that the system evolves rapidly into a single-field system. We have checked numerically the validity of the $\alpha X\ll1$ approximation for the values of $\hat{\a}$ and $\hat{\alpha}'$ we chosen and found that $\alpha X$ never exceeds $\mathcal{O}(10^{-3})$.

Inserting the equation of the minimum line ({\ref{MIN}}) into the action ({\ref{ACT-FINAL}}) we obtain a single-field action with the potential reduced to just the Palatini-${\cal{R}}^2$ term and the kinetic function augmented by the extra contribution of the $\sigma$-kinetic term $(\nabla\sigma_0)^2=(\nabla\phi)^2(\sigma_0'(\phi)\,)^2$. We get
\be {\cal{S}}=\int\,{\rm d}^4x\,\sqrt{-g}\Bigg\{\frac{1}{2}R(g)-\frac{1}{2}\bar{\cal{K}}(\phi)(\nabla_\m\phi)^2-\bar{U}(\phi)\Bigg\}\,,{\label{ACT-PHI}}\ee
where
 \be 
\begin{split}
    &\bar{\mathcal{K}}(\phi)=\bar{\mathcal{K}}_1(\phi)+\bar{\mathcal{K}}_2(\phi)\Bigg(-\frac{(f(\phi ) g(\phi )) \left(2 f(\phi ) f'(\phi )+4 \beta  V'(\phi )\right)}{2 \left(f^2(\phi )+4 \beta  V(\phi )\right)^2}\\&+\frac{g(\phi ) f'(\phi )}{2 \left(f^2(\phi )+4 \beta  V(\phi )\right)}+\frac{f(\phi ) g'(\phi )}{2 \left(f^2(\phi )+4 \beta  V(\phi )\right)}   \Bigg)^2\\\\&
    \bar{\mathcal{K}}_1(\phi)=\frac{f(\phi)K(\phi)-2\hat{\a}V(\phi)}{f^2(\phi)+4\b V(\phi)}\,,\,\,\,\,\,\,\,\,\,\,\,
    \bar{\mathcal{K}}_2(\phi)=\frac{24 \left(f^2(\phi )+4 \beta  V(\phi )\right)^2}{4 f^2(\phi ) g^2(\phi )+\left(f^2(\phi )+4 \beta  V(\phi )\right)^2}\\\\&
    \bar{U}(\phi)=\frac{V(\phi)}{f^2(\phi)+4\beta V(\phi)}
\end{split}
\ee 
Note that, although there is no trace of the Holst coupling parametric function in the single-field potential, there is a strong dependence on it, as well as on the derivative coupling parameter $\hat{\alpha}$, in the kinetic function $\bar{\cal{K}}(\phi)$.
\section{Inflation}
\label{inflation}
We proceed to analyze the inflationary profile of the single-field model assuming the slow-roll approximation. We start with the scalar (${\cal{P}}_{\zeta}$) and tensor (${\cal{P}}_T$) power spectra selecting an pivot scale $k_{\star}$ that exited the horizon. Their expressions are 
\be
\label{eq:spectra}
\mathcal{P}_\zeta (k)=A_s \left(\frac{k}{k_\star} \right)^{n_s -1},\,\,\,\,{\textit{where}}\,\,\, \,\, A_s\simeq\frac{1}{24\pi^2}\frac{\bar{U}(\phi_\star)}{\epsilon_{\bar{U}}(\phi_\star)},\,\,\,\,\,\,{\textit{and}}\,\,\,\,\,\,\, \mathcal{P}_T (k)\simeq\frac{2\bar{U}(\phi_\star)}{3\pi^2} \left(\frac{k}{k_\star} \right)^{n_t}\,,
\ee
$A_s$ being the amplitude of the power spectrum of scalar perturbations.
The scalar ($n_s$) and tensor ($n_t$) spectral indices given by
\be
\label{eq:index}
n_s-1=\frac{{\rm d} \ln \mathcal{P}_\zeta (k) }{{\rm d} \ln k} \simeq -6\epsilon_{\bar{U}} +2\eta_{\bar{U}}  \qquad \text{and} \qquad n_t= \frac{{\rm d} \ln \mathcal{P}_T (k) }{{\rm d} \ln k}\,,
\ee
characterize the scale-dependence of the power spectra~\eqref{eq:spectra}. The tensor-to-scalar ratio is defined as\be
\label{eq:ttsr}
    r= \frac{\mathcal{P}_T (k)}{\mathcal{P}_\zeta (k)} \simeq 16\epsilon_{\bar{U}}.
\ee
These equations are expressed in terms of the potential slow-roll parameters 
\be
\label{eq:pslp}
\eps_{\bar{U}} = \frac{1}{2\bar{\mathcal{K}}(\phi)} \left( \frac{\bar{U}'(\phi)}{\bar{U}(\phi)} \right)^2 \qquad \text{and} \qquad \eta_{\bar{U}} = \frac{\left(\bar{\mathcal{K}}^{-1/2}(\phi) \bar{U}'(\phi)\right)'}{\bar{\mathcal{K}}^{1/2}(\phi)\bar{U}(\phi)}\,.
\ee
The primes denote derivatives with respect the scalar field, while
both slow-roll parameters are assumed to be small $(\ll 1)$ during inflation. 
The duration of inflation is measured by the number of $e$-folds, given by
\be
\label{eq:efolds}
N_\star =\int^{\phi_{\star}}_{\phi_{\rm end}} \bar{\mathcal{K}}(\phi) \frac{\bar{U}(\phi)}{\bar{U}'(\phi)}{\rm d}\phi\,,
\ee
where the end of inflation at $\phi_{end}$ is determined by $ \epsilon_{\bar{U}}(\phi_{end})\,\approx\,1\,$.
We specify the model taking the scalar field $\phi$ to have a canonical kinetic term ($K(\phi)=1$) and a quartic potential
\begin{table}[h!]
    \centering
    \begin{tabular}{||c|c|c||}
        \hline \hline
        $N_\star$ (e-folds) & $\phi_\star$ ($\hat{\a}=-10^{3}$) & $\phi_\star$ ($\hat{\a}=-10^{10}$) \\
        \hline
        50  & 19.55 &  6.15\\
        55  & 20.54 &  6.30\\
        60  & 21.48 &  6.44 \\ 
        \hline
        $N_\star$ (e-folds) & $\phi_\star$ ($\hat{\a}=-10^{3}$) & $\phi_\star$ ($\hat{\a}=-10^{10}$) \\
        \hline
        50  & 12.09 &  5.85\\
        55  & 12.72 &  6.01\\
        60  & 13.33 &  6.16 \\ 
        \hline \hline
    \end{tabular}
    \caption{The first table corresponds to $\xi=1$, $\bar{\xi}=0$, $\bar{\xi}'=1$, $\b=10^6$ while the second to $\xi=1$, $\bar{\xi}=1$, $\bar{\xi}'=0$, $\b=10^6$.}
    \label{table1}
\end{table}

\be 
V(\phi)=\frac{\l}{4}\phi^4\,.\ee 
The non-minimal coupling functions are taken to be
\be f(\phi)=1+\xi\phi^2,\quad g(\phi)=\bar{\xi}\phi+\bar{\xi}'\phi^3\,.
\ee
The number of e-folds $N_{\star}$ as a function of the initial value of the field $\phi_{\star}$ is obtained by numerical integration of ({\ref{eq:efolds}}). Characteristic values are given in table~\ref{table1} for representative values of the parameters. 

\begin{figure}[t!]
  \centering
  \begin{subfigure}[b]{0.48\textwidth}
    \includegraphics[width=\textwidth]{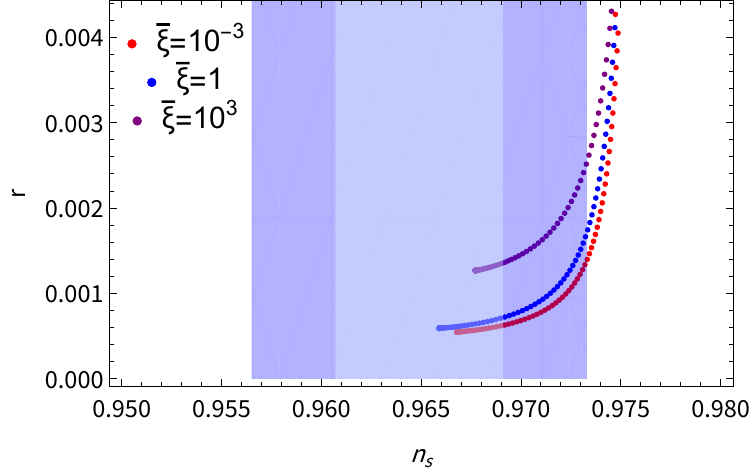} 
    \caption{}
    \label{subfig3:a}
  \end{subfigure}
  \hfill
  \begin{subfigure}[b]{0.48\textwidth}
    \includegraphics[width=\textwidth]{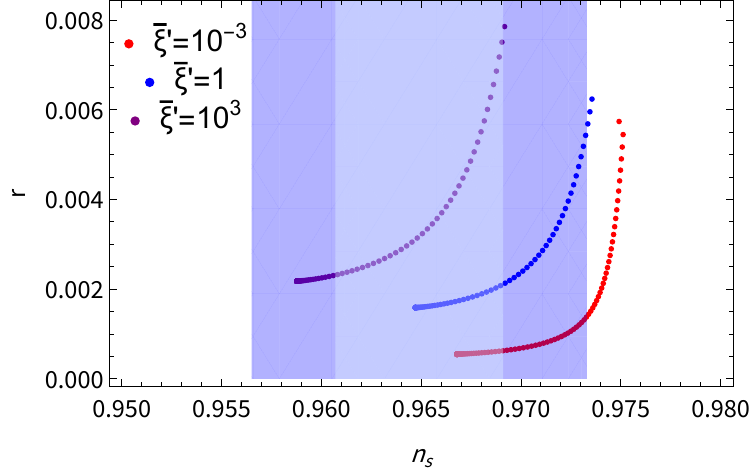}
    \caption{}
    \label{subfig3:b}
  \end{subfigure}
  \caption{The spectral index $n_s$ with respect to $r$ in order to have $N_\star=60$. The parameters we chosen are $\b=10^6$ and $\l$ is fixed in order to have $A_s\sim 2.1\cdot 10^{-9}$. (a) $\xi=1$, $\bar{\xi}'=0$ and $\bar{\xi}=10^{-3}$ (red), $\bar{\xi}=1$ (blue) and $\bar{\xi}=10^3$ (purple). (b) $\bar{\xi}'=10^{-3}$ (red), $\bar{\xi}'=10$ (blue), $\bar{\xi}'=10^3$ (purple). The lines trace the different values of the derivative coupling parameter $\hat{\alpha}$ with a significant effect on observables for values $10^2\leq|{\hat{\a}}|\leq10^{10}$ for which $n_s$ increases with $\hat{\a}$. We have indicated with light-blue the $1\s$ and with dark blue the $2\s$~\cite{Planck:2018jri,BICEP:2021xfz}.}
  \label{SPECT-1}
\end{figure}

The existing observations of CMB~\cite{Planck:2018jri,BICEP:2021xfz}, corresponding to the Planck, BICER/Keck and BAO data, constrain the observables at the pivot scale $k_{\star}=0.05\,{\rm Mpc}^{-1}$ to be 
\be A_s=(2.10\pm 0.03)\times 10^{-9},\,\,\,n_s=0.9649\pm 0.0042\,\,(1\sigma),\,\,\,\,r<0.03\,\,.\ee

\begin{figure}[t!]
  \centering
  \begin{subfigure}[b]{0.48\textwidth}
    \includegraphics[width=\textwidth]{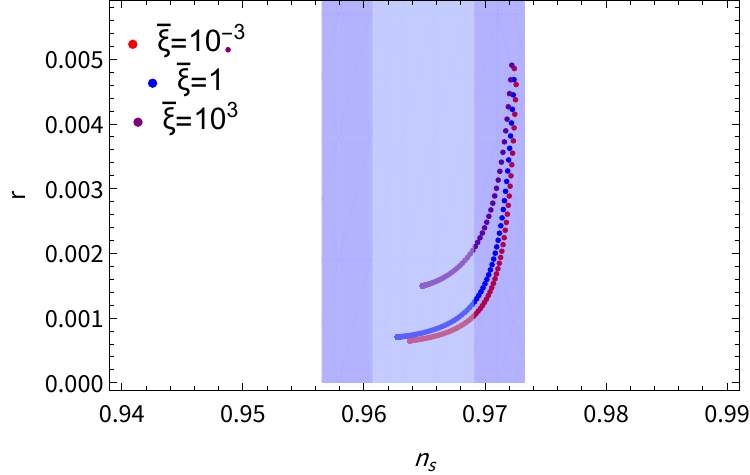} 
    \caption{}
    \label{subfig4:a}
  \end{subfigure}
  \hfill
  \begin{subfigure}[b]{0.48\textwidth}
    \includegraphics[width=\textwidth]{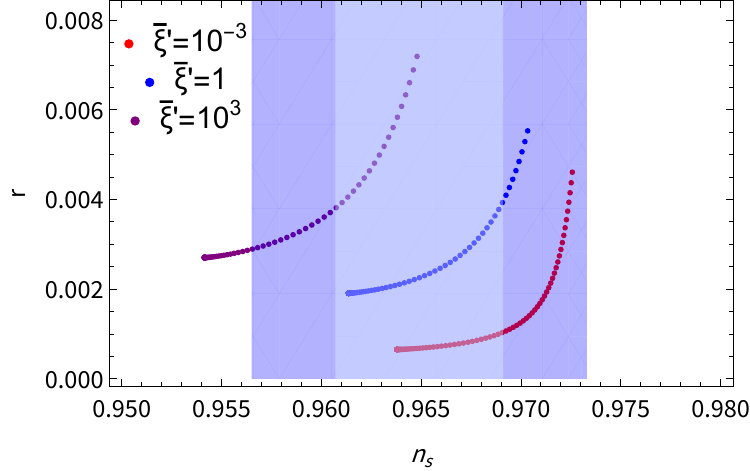}
    \caption{}
    \label{subfig4:b}
  \end{subfigure}
  \caption{The spectral index $n_s$ with respect to $r$ in order to have $N_\star=55$. The parameters we chosen are $\b=10^6$ and $\l$ is fixed in order to have $A_s\sim 2.1\cdot 10^{-9}$. (a) $\xi=1$, $\bar{\xi}'=0$ and $\bar{\xi}=10^{-3}$ (red), $\bar{\xi}=1$ (blue) and $\bar{\xi}=10^3$ (purple). (b) $\xi=1$ and $\bar{\xi}=0$, $\bar{\xi}'=10^{-3}$ (red), $\bar{\xi}'=1$ (blue), $\bar{\xi}'=10^3$ (purple). The lines trace the different values of the derivative coupling parameter $\hat{\alpha}$ with a significant effect on observables for values $10^2\leq|{\hat{\a}}|\leq10^{10}$ for which $n_s$ increases with $\hat{\a}$. We have indicated with light-blue the $1\s$ and with dark blue the $2\s$~\cite{Planck:2018jri,BICEP:2021xfz}.}
  \label{SPECT-2}
\end{figure}

In Fig.~\ref{SPECT-1} and~\ref{SPECT-2} we have plotted $n_s$ versus $r$ for various choices of the coupling parameters for $N_\star\sim 55$ and $N_\star\sim 60$. The value of the quartic coupling $\lambda$ is determined from the above $A_s$ value. For the representative parameter values of table~\ref{table1} it has a negligible dependence on $\hat{\alpha}$, as long as $|\hat{\alpha}|<10^{10}$. {{ As we have mentioned above the invertibility of the disformal transformation requires that $\hat{\a}<0$}}. As it can be seen in Fig.~\ref{SPECT-1} the calculated values of the indices $n_s$ and $r$ lie for the most part within the acceptable range of observations. Increasing values of $\hat{\alpha}$ lead to an increasing tensor-to-scalar ratio $r$ and tend to push the model values outside the acceptable domain. This is more evident in Fig.~\ref{subfig3:a} which corresponds to a linear Holst coupling. As it is more evident in Fig.~\ref{subfig3:b}, which refers to a cubic Holst coupling, larger values of the corresponding parameter are favored, although they also lead to a faster increase of $r$ with $\hat{\alpha}$. In these cases we found numerically that for $|\hat{\a}|\sim 10^9$ the scalar tilt becomes $n_s\sim0.9733$. Therefore, for all values of $|\hat{\a}|>10^9$ the model is pushed outside of the observational data. The known inflationary behavior in the absence of derivative couplings, due to the $R^2$ term and/or the $\xi$-coupling is sustained as long as $\hat{\alpha}$ does not exceed certain value, depending on the form and strength of the Holst coupling. We note that in $\hat{\a}=0$ limit the values of the observables are in agreement with the previous study~\cite{Gialamas:2022xtt}. In the case where the contribution from the Holst term is set to zero, the predictions of the model coincide with those of~\cite{Gialamas:2024jeb}. We have numerically verified that our results are identical to the previous ones. There it was established that $n_s$ and $r$ tend to increase for increasing $\bar{\xi}$ and $\bar{\xi}'$. This feature remains also true in the present case as it can be seen in the corresponding figures.

 In the large $\phi$ limit the kinetic function is of the form $\bar{K}(\phi)\approx |\hat{\alpha}|  \lambda /2 \left(\beta  \lambda +\xi ^2\right)$. Then, the corresponding canonical field $\phi_c=\int\,d\phi\sqrt{K(\phi)}$ is 
$\phi_c\approx\,\phi \left(\frac{|\hat{\alpha} | \lambda }{2(\beta  \lambda +\xi ^2)}\right)^{1/2}$,
which can easily lie in the subplanckian regime for the representative values of parameters adopted above.

\section{Summary}
\label{summary}
In the present article we have considered the framework of Einstein-Cartan gravity and studied a model of a non-minimally coupled scalar with a quartic $\phi^4$ potential that includes up to quadratic terms of the Ricci scalar and the Holst invariant. The found that the model is characterized by an extra dynamical pseudoscalar degree of freedom $\sigma$, associated with the presence of the Holst term. We have analyzed the evolution of the two-field system in an FRW background and found that it evolves rapidly along a trajectory of minimal potential, being reduced into an effective single-field system. We proceeded to study slow-roll inflation for the single-field system, analyzing the effect of the derivative couplings on its inflationary behavior. We focused on the determination of the observables $n_s$ and $r$ for representative values of the model parameters and studied their dependence on the derivative coupling parameter $\hat{\alpha}$, finding that an increase in $\alpha$ results in an increase of their corresponding  values. We find that the model is cosistent with the latest  combined observational data from Planck, BICEP/Keck and BAO for relativelly wide range of its parameters. Nevertheless, for large derivative coupling parameter values $|\hat{\a}|>10^9$, the model becomes incosistent with observations. It should be noted that the ${\cal{R}}^2$ and/or the non-minimal $\xi$-coupling are central to the consistent inflationary behavior of $\phi^4$, which is sustained as long as the derivative coupling stays below these values.

\acknowledgments

We aknowledge helpfull discussions with Ioannis Gialamas.

\appendix

\section{Effective Metric Action} 
\label{appendix}
The algebraic solution to the equations of the torsion vectors, in terms of $E_{ \mu}$, given in ({\ref{EPSILON}}), is
  \be\begin{split}
&\hat{T}_\m=\frac{1}{A}E_\m-\frac{B}{A(A+BX)}X_{\m\n}E^\n\\
\\&
T_{\m}=\frac{1}{A}\Big(F_4+\frac{1}{6}F_5X \Big)E_\m+\Bigg\{\frac{1}{3A}F_5-\frac{B}{A(A+BX)}\Big(F_4+\frac{1}{6}F_5X\Big)-\frac{1}{3}F_5\frac{BX}{A(A+BX)}\Bigg\}X_{\m\n}E^\n,
\end{split}
\ee
where 
\be A=\frac{1}{16}+\Big(F_4+\frac{1}{6}F_5X\Big)^2\,\,\,\,{\text{and}}\,\,\,\,
    B=\frac{1}{3}\Big\{2F_5\Big(F_4+\frac{1}{6}F_5X\Big)+\frac{1}{3}F_5^2X\Big\}
\ee
Substituting back into the action ({\ref{ACT-3}}) we obtain

\be 
\begin{split}
   \mathcal{S}&=\int {\rm d}^4x\sqrt{-g}\Bigg\{\frac{1}{2}R-\frac{\Delta^2}{3}\Big[F^2E^2+H(2F+HX)(E\cdot X\cdot E)\Big]+\frac{\Delta^2}{48}\Big[E^2\\&+G(GX+2)(E\cdot X\cdot E)\Big]+\frac{2}{3}F_4\Delta^2\Big[FE^2+(FG+H+HGX)(E\cdot X\cdot E)\Big]\\&+\Delta\Big[E\cdot\nabla F_4+GE\cdot X\cdot\nabla F_4)\Big]+\frac{\Delta}{3}\Big[E\cdot \nabla(F_5X)+G(E\cdot X\cdot\nabla (F_5X))\Big]\\&-\frac{\Delta}{3}\Big[ \nabla_\m(\tns{X}{^\m_\n}F_5)E^\n+G\nabla_\m(\tns{X}{^\m_\n}F_5)\tns{X}{^\n_\r}E^\r \Big]+\frac{2}{9}\Delta^2F_5\Big[(F+FGX\\&+HX+HGX^2)(E\cdot X\cdot E)\Big]+\frac{\Delta^2}{9}F_5X\Big[FE^2+(FG+HGX+H)(E\cdot X\cdot E) \Big]\\&-\frac{1}{2}\g_1(1+\g X)^{-1/2}K(\phi)X-\g_1^2(1+\g X)^{1/2}U    \Bigg\} 
\end{split}
{\label{ACT-4}}
\ee
expressed in terms of the functions
\be
\Delta=A^{-1},\,\,\,
F=F_4+\frac{1}{6}F_5X,\,\,\,
G=-\frac{B}{A+BX}\,,\,\,\,\,
H=\frac{1}{3}F_5+G\Big(F+\frac{1}{3}F_5X\Big)
\ee
In order to simplify notation, from now on we shall denote $F_4=\sigma$. Then, $F_5=\gamma F_4=\gamma\sigma$.
The scalar field action in ({\ref{ACT-4}}) consists of two parts, namely a part 

\be
\begin{split}
   \mathcal{S}_I&=\int {\rm d}^4x\sqrt{-g}\Bigg\{-\frac{\Delta^2}{3}\Big[F^2E^2+H(2F+HX)(E\cdot X\cdot E)\Big]+\frac{\Delta^2}{48}\Big[E^2\\&
   +G(GX+2)(E\cdot X\cdot E)\Big]+\frac{2}{3}\sigma\Delta^2\Big[FE^2+(FG+H+HGX)(E\cdot X\cdot E)\Big]\\&
   +\Delta\Big[E\cdot\nabla \sigma+GE\cdot X\cdot\nabla \sigma)\Big]+\frac{\Delta}{3}\Big[E\cdot \nabla(F_5X)+G(E\cdot X\cdot\nabla (\gamma\sigma X))\Big]\\&
   -\frac{\Delta}{3}\Big[ \nabla_\m(\tns{X}{^\m_\n}\gamma\sigma)E^\n+G\nabla_\m(\tns{X}{^\m_\n}\gamma\sigma)\tns{X}{^\n_\r}E^\r \Big]+\frac{2}{9}\Delta^2\gamma\sigma\Big[(F+FGX\\&
   +HX+HGX^2)(E\cdot X\cdot E)\Big]+\frac{\Delta^2}{9}\gamma\sigma X\Big[FE^2+(FG+HGX+H)(E\cdot X\cdot E) \Big]   \Bigg\} 
\end{split}
{\label{ACT-5}}
\ee
and a part
\be {\cal{S}}_{II}=\int\,{\rm d}^4x\,\sqrt{-g}\Bigg\{-\frac{1}{2}\g_1(1+\g X)^{-1/2}K(\phi)X-\g_1^2(1+\g X)^{1/2}U \Bigg\} {\label{ACT-6}}\ee
${\cal{S}}_I$ contains only kinetic terms. To $\mathcal{O}(X)$ it is a function of $\sigma$, its derivative and derivatives of $\phi$, namely
\be
\begin{split}
\mathcal{S}_I&=\int\,{\rm d}^4x\,\sqrt{-g}\Bigg\{\frac{-12+64(\alpha_2X-3)\sigma^2}{(1+16\sigma^2)^2}(\nabla_\m\sigma)^2-\frac{384\sigma^2}{(1+16\sigma^2)^2}X^{ \mu\nu}\nabla_{ \mu}\sigma\nabla_{ \nu}\sigma\\&
+\frac{(-8+\sigma+32\sigma^2(-20+\sigma+8\sigma^2))\alpha_2}{(1+16\sigma^2)^2}\Big(\nabla_{ \mu}\sigma\nabla^{ \mu}(\sigma X)-\nabla_{ \mu}(X^{ \mu\nu}\sigma)\nabla_{ \nu}\sigma\Big)\,\,\Bigg\}
\end{split}
\label{ACT-7}
\ee
The part ${\cal{S}}_{II}$ is a function of all the scalars, including the auxiliary $\chi$.

\bibliography{Einstein_Cartan}{}
\end{document}